\documentclass[12pt,preprint,amsmath,amssymb,nofootinbib]{revtex4}

 \usepackage{epsfig}

\usepackage{amsmath,amsfonts}

\newcommand{\CO}{{\cal O}}

\newcommand{\ket}{{\rangle}}

\newcommand{\vev}[1]{{\langle #1 \rangle}}

\newcommand{\tr}{\hbox{ Tr}}

\begin{document}

\title{Numerical tests of AdS/CFT at strong coupling}

\author{David Berenstein$^{1,2}$, Randel Cotta $^{ 3}$, Rodrigo Leonardi $^1$
}
\affiliation{$^1$ Department of Physics, University of California at Santa Barbara, CA 93106\\
$^2$ Isaac Newton Institute for mathematical Sciences, Cambridge CB3 OEH, UK
\\
$^3$ Department of Physics, Stanford University, Stanford, CA 94305  }

\preprint{NI07082}

\begin{abstract}
We study various correlation functions (two and three point functions) in a large $N$ matrix model of six commuting matrices with a numerical Monte Carlo algorithm.  This is equivalent to a model of a gas of particles in six dimensions with a confining quadratic potential and logarithmic repulsions at finite temperature, where we are measuring the leading order non-gaussianities in the thermal fluctuations.
 This is a simplified model of the low energy dynamics of ${\cal N}=4 $ SYM at strong coupling. We find strong evidence that the simplified matrix model matches with the dual gravitational description of  three point functions in the AdS/CFT correspondence.
\end{abstract}

\maketitle

\section{Introduction}

The AdS/CFT correspondence \cite{M} in its simplest setting states that an ordinary quantum field theory in $d=4$ dimensions, the maximally supersymmetric Yang Mills theory with gauge group $U(N)$, is equivalent to type IIB string theory as a theory of gravity compactified on an
$AdS_5\times S^5$ geometry. As such, this equivalence provides a way to answer questions about the nature of quantum gravity by performing  calculations in the dual quantum field theory.

The radius $R$ of the $S^5$ geometry in string units is related to the 't Hooft coupling of the field theory $\lambda= g^2 N$ via a scaling $R^4 \sim \lambda$. The $S^5$ sphere is large in string theory units if $R>>1$. In this regime, it is expected that semiclassical calculations in supergravity are a reliable description of the physics. In other regimes, it is not known how to do    calculations in the gravity side of the correspondence because one would need to understand how stringy corrections affect the dynamics of gravity.
Thus, in order to compare the field theory and quantum gravity one should expand the field theory quantities at large values of $\lambda$. This is, one should analyze the field theory at strong coupling.
 
 On general grounds studying field theory dynamics at strong coupling is a hard problem. 
In lattice gauge theory occasionally there are ways to address this problem \cite{KGW}, although since the ${\cal N}=4 $ SYM theory does not confine (it is a non-trivial conformal field theory) a lattice definition of the field theory might not be very useful and it might be hard to extract information from such a formulation.

A proposal for how to do a strong coupling expansion of the ${\cal N}=4 $ SYM compactified on a $S^3$ was proposed in \cite{BlargeN}. The compactification on $S^3$ provides an infrared regulator of the dynamics. Morever, via the operator state correspondence, the spectrum of the Hamiltonian of the field theory on the $S^3$ is equivalent to the spectrum of dimensions of local operator insertions in Euclidean space. Thus, one can address the problem of computing operator dimensions and correlation functions by using Hamiltonian methods.

In particular, in the proposal of \cite{BlargeN} an expansion of the action of all fields of the ${\cal N}=4$ in spherical harmonics on the $S^3$ was truncated to the s-wave of the scalar
field modes in the field theory. One also needs to include the s-wave of the $A_0$ component of the gauge field to be able to impose the gauge constraint. The intuition that led to this truncation was that supersymmetric states in the free field limit only excite these modes and that
supersymmetry should account  for cancelations of quantum corrections. Thus a naive truncation might be good enough to describe the dynamics, even at strong coupling.

The effective Hamiltonian for this setup is a matrix quantum mechanics of six hermitian $N\times N$ matrices $\vec X$, and their momenta
\begin{equation}
H = \frac 12 \tr \vec P^2+\frac 12\tr \vec X^2+ \frac 14 A\sum_{i,j}\tr( [X_i,X_j]^2) 
\end{equation}
where the coupling constant $A$ has been calculated in \cite{BCV}. It is proportional to the gauge coupling constant
squared. This has to be supplemented by the gauge constraints.  At large $N$, one expects that the eigenvalues of $X$ are of order $\sqrt N$. The first two terms of the Hamiltonian would be of order $N^2$, but the last term, the commutator squared term, would be of order $g^2N N^2= \lambda N^2$, so for
general configurations of matrices of this size the potential term is much larger than the kinetic term. Under these circumstances one expects that one should expand the system around configurations that minimize the large potential term. This is, one should expand around configurations of commuting matrices. The expansion that was proposed in \cite{BlargeN} is exactly to solve the reduced (gauged) matrix quantum mechanics model of commuting matrices and to treat all other modes perturbatively around this (non-perturbative) background. 
Commuting matrices can be diagonalized simultaneously by a gauge transformation, so that only the eigenvalues are dynamical variables. For each diagonal component there are 6 coordinates, the eigenvalues of each matrix (we will label them as $\vec x_i$).

 It was found that
the wave function of the ground state for this reduced set of degrees of freedom was a Gaussian
\begin{equation}
\psi_0 = \exp( -\frac 12\sum_i \vec x_i^2)
\end{equation}
but that there is also a measure term  that affects the calculation of averages in the quantum problem. This measure is 
\begin{equation}
\mu^2 = \prod_{i<j}|\vec x_i-\vec x_j|^2
\end{equation}
a generalization of the Van der Monde determinant.

Quantum averages of operators that are restricted to this set of diagonal variables are computed by the following integrals
\begin{equation}
\vev{\CO}= \frac{\int \mu^2 \exp(-\sum_i \vec x_i^2) \CO(x)}{\int \mu^2 
\exp(-\sum_i \vec x_i^2) }
= \frac{\int \exp(-\sum_i \vec x_i^2- \sum_{i<j} \log|\vec x_i-\vec x_j|^2) \CO(x)}{\int \mu^2 \exp(-\sum_i \vec x_i^2) }\label{eq:vev}
\end{equation}

This is equivalent to studying thermal correlations for a Boltzman gas of $N$ particles in six dimensions in the presence of a one body potential ($\vec x^2$), and with a two body logarithmic repulsive interaction ($\log|\vec x-\vec y|^2)$). This is a somewhat unusual problem in statistical mechanics, but it can be approached numerically. This approach was initiated in \cite{BCott}, where very few quantities were computed, confirming some of the theoretical ansatz \cite{BlargeN, BCV} that solved the saddle point approximation for the thermodynamic limit of the distribution (see also \cite{GHHK} for some related problems in the thermal field theory case). In particular, we expect that in the thermodynamic limit the theory will be described by some density of particles $\rho(x)$ and thermal fluctuations of the density.

In this paper, the numerical study of this ensemble is continued. In particular we compute 
the density fluctuations of various modes of the ensemble for various values of $N$ and we study the approach of the fluctuations to the thermodynamic limit. We also compute the leading order non-gaussianities of the density fluctuations. These are related to three point (extremal) correlation functions in the conformal field theory. These have been studied in the general case \cite{LMRS} where a non-renormalization theorem for BPS operators was conjectured. However, it was found that in the extremal case the computation requires solving some subtleties \cite{D'H5} and once this is done, the results of \cite{LMRS} 
for supergravity can be used. A non-renormalization theorem for these extremal correlators
was proved in \cite{EHSSW} using the techniques of harmonic superspace. 

The  point of view we will take in this paper is that we want to verify if the approximations that 
were taken to study the dynamics of ${\cal N}=4 $ at strong coupling in \cite{BlargeN} are a complete description of the low energy dynamics that leads to supergravity: we do not need the other modes in the field theory to reproduce the results expected by the non-renormalization theorem at strong coupling, and we can compare with the results found using supergravity in \cite{LMRS,D'H5}. 

The paper is organized as follows. In section two we explain how the 
two and three point functions are calculated in the free field theory limit of ${\cal N}=4$ 
SYM. In particular, we show how the OPE coefficients are computed by a free matrix model. 
We also discuss in this section how this calculations was matched to supergravity, suggesting a non-renormalization theorem for three point functions. Next, in section \ref{sec:ssc} we give 
a description of how we set up the calculations of the three point functions for strong coupling
within the wave function approach. In section \ref{sec:nr} we give a detailed presentation of our numerical results for two and three point functions at strong coupling. Then we conclude. 
We have also included an appendix where we describe how the statistical error bars were calculated and some details of how the Monte-Carlo code was calibrated \footnote{
The computer code used to generate the data with intstructions for compilation is available on request from D. B.}.

\section{Free matrix model results}\label{sec:fmm}

Conformal field theories are usually determined by the two and three point functions of primary operators (these are given by local insertions of composite operator $\CO_i(x)$). 
The two point functions (of scalar operators) are given by 
\begin{equation}
\vev{ \CO_i(x) \CO_j(y)} \sim \frac{C_{ij} \delta_{\Delta_i, \Delta_j}}{|x-y|^{ \Delta_i+\Delta_j}}
\end{equation}
where $\Delta_{i,j}$ are the dimensions of the corresponding operators, while $C_{ij}$ is a set of numbers (symmetric in $i,j$)
and the three point functions of scalar operators are given by
\begin{equation}
\vev{ \CO_i(x) \CO_j(y) \CO_k(z)} \sim \frac{C_{ijk}} {|x-y|^{ \Delta_i+\Delta_j-\Delta_k}|x-z|^{ \Delta_i+\Delta_k-\Delta_j}|y-z|^{ \Delta_k+\Delta_j-\Delta_i}}
\end{equation}
where the $C_{ijk}$ are structure constants of the theory.

The matrix $C_{ij}$ is called the Zamolodchikov metric, and it is a standard procedure to diagonalize the metric so that $C_{ij}= \delta_{i,j}$. In this sense the exact normalization of the two point functions is somewhat unphysical, unless there are degeneracies in the list of operators with given quantum numbers.

The coefficients $C_{ijk}$ are also the coefficients of the Operator product expansion of
$\CO_{i}$ and $\CO_{j}$ into $\CO_{k}$ (or any permutation of the three).
\begin{equation}
\CO_i(x) \CO_j(0) \sim \sum_k \frac{C_{ijk}}{|x|^{\Delta_i+\Delta_j-\Delta_k}} \CO_{k}(x)
\end{equation}
This is usually an asymptotic series, organized starting from the most singular terms 9those with smallest value of $\Delta_k$.

In the special case where for the leading order term $\Delta_k = \Delta_i+\Delta_j$, the operator product expansion is non-singular, and one can take the limit  $x\to 0$ in the above formula. This makes it possible to have a composite operator $(\CO_i\CO_j)(x)$.

In the special case of ${\cal N}=4$ SYM, there is a special class of operators of protected dimension. This is, the dimension of the operator when calculated in the free field theory
limit is identical to the dimension of the operator in the interacting theory. 

The simplest such operators are given by symmetric traceless tensors of $SO(6)$, 
\begin{equation}
\CO_{i_1, \dots, i_k}  \sim  \tr( X_{i_1}\dots X_{I_k}) (0)
\end{equation}
where $\delta^{ij}\CO_{i,j, \dots, i_k}=0$ and 
$\CO_{i_1, \dots i_s\dots i_m \dots i_k}= \CO_{i_1, \dots i_m\dots i_s, \dots, i_k}$. This is a composite field of $k$ free fields, and it is of dimension $k$. 
These were first described in the AdS/CFT setup in \cite{W}, where they were matched with supergravity fluctuations \cite{KRN}. Each trace in interpreted as a single graviton state. 

Each of these single trace operators can be related to a simpler one $\tr(Z^k)$, where we choose a particular complex combination of the fundamental scalar fields $Z= X_1+iX_2$. This 
can be thought of as the highest weight state of the associated $SO(6)$ representation.

The (free field) two point functions of these operators depend on how we choose to normalize the fields in the field theory action. 

In free field theory, to leading order in planar diagrams, one finds that 
\begin{equation}
\vev{\tr(Z^k)(x) \tr(\bar Z^k)} \sim \frac{k N^k}{|x|^{2k}}( 1+O(1/N^2))= \frac{C_{k,k}}{|x|^{2k}}\label{eq:twoptfree}
\end{equation}
More precisely, one can compute $C_{k,k}$ by performing integrals in a Gaussian matrix model for a complex matrix $z$ \cite{CJR},
\begin{equation}
C_{k,k}= \frac{\int (dz d\bar z)^{N^2}
\exp( - \tr(z\bar z)) \tr(z^k)\tr(\bar z^k) }{\int (dz d\bar z)^{N^2}
\exp( - \tr(z\bar z)) }\sim \vev{ \tr(z^n)\tr(\bar z^n)}\label{eq:gaussian1}
\end{equation}
where the average in the right hand side is understood as a matrix model integrals (overlap). 

One also finds that multi-traces of $Z$ are protected operators, and they are different than single trace operators. Thus one has the problem that
there is more than one operator of a given dimension for a given set of quantum numbers. 
In general the operators mix and this problem needs to be resolved. A complete solution in the free field limit was given in \cite{CJR}, where a complete orthogonal basis was found.
The general problem of multitrace mixing can also be understood in terms of the complex matrix model, by calculating averages  of the form
\begin{eqnarray}
C_{k_1, \dots k_s; \tilde k_1, \dots, \tilde k_t }&=& \frac{\int (dz d\bar z)^{N^2}
\exp( - \tr(z\bar z)) \prod_j \tr(z^{k_j})\prod_{j'}\tr(\bar z^{\tilde k_j'}) }{\int (dz d\bar z)^{N^2}
\exp( - \tr(z\bar z)) }\label{eq:gaussian2}\\
&=& \vev{ \prod_j \tr(z^{k_j})\prod_{j'}\tr(\bar z^{\tilde k_j'})}
\end{eqnarray}
If $k_1+\dots+k_s$ is held fixed, the numbers $C$ provide a non-diagonal 
metric on the space of half BPS states.

Of particular interest to us is the correlation function $C_{n, m-n; m}$, and to leading
order this is given by 
\begin{equation}
C_{n, m-n; m} = n(m-n) m N^{m-1} ( 1+ O(1/N^2))
\end{equation}
in the same normalization for the fields where the two point functions
are calculated by 
\begin{equation}
C_{k,k} = k N^k ( 1+ O(1/N^2))
\end{equation}
$C_{n, m-n,;m}$ can be interpreted both as a two point function for describing mixing between the operators $\tr(Z^n)\tr(Z^{m-n}$ and $\tr(\bar Z^m)$, and as a three point function
for extremal correlators.

To study the physically relevant information encoded in these two and three point functions it 
is natural to calculated them in a case where the two point functions are normalized relative to the Zamolodchikoiv metric for two point functions \cite{LMRS}. Thus, we would compute the 
normalized correlator
\begin{equation}
N_{n, m-n; m} = \frac{C_{n, m-n; m }}{\sqrt{ C_{n,n} C_{m-n, m-n} C_{m,m}}}
\sim \frac{\sqrt{n(m-n) m }}{N} ( 1+ O(1/N^2)) \label{eq:compare}
\end{equation}

The non-renormalization conjecture of \cite{LMRS} is that $N_{n,m; m-n}$ does not change in value in extrapolating from free field theory to strong coupling. This is the set of numbers we will compute at strong coupling with the proposal \cite{BlargeN}.

The other useful calculation to do is the following
\begin{equation}
C_{k, k, \dots k; k, k \dots k}\sim C_{k,k}^s s! (1+O(1/N^2) ) \label{eq:gauss}
\end{equation}
where we have take $s$ copies of the same trace. It is easy to show by planar counting 
that the right hand side in the above equation has a factor of $s!$. This is because the leading contractions are from disconnected diagrams between the different groups of traces. The factor of $s!$ just counts how the different permutations of pairings of the traces.

The equation (\ref{eq:gauss}) states that to leading order in $1/N$,  $\tr(z^k)$ can be interpreted as a raising operator in a harmonic oscillator $\tr(z^k) \sim \alpha_k^\dagger$. This is because
\begin{equation}
||(\alpha_k^\dagger)^s | 0 \ket ||^2 \sim s!
\end{equation}
thus if we think of equation (\ref{eq:gaussian1}) as a statistical ensemble, we should find that
the distribution of $\tr(z^k)$ is Gaussian. This is familiar from quantum mechanics for a harmonic oscillator, where the ground state wave function is
\begin{equation}
\psi_0(x) \sim \exp( -x^2/2 \sigma)
\end{equation}
and we find that the distribution of $x$ in the ground state is Gaussian. This also applies to
the combination $\alpha^\dagger \sim x+ i p$, where we now get a Gaussian distribution in the complex plane (phase space for a single variable).

One point functions of powers of a complex matrix vanish
\begin{equation}
\vev{\tr(z^n)} = 0
\end{equation}
for $n>0$. This is because the gaussian measure is invariant under phase shifts $z\to \exp(i\theta) z $. This property is inherited by all correlation functions, so correlation functions that are not invariant under that shift will automatically vanish.

\section{Strong coupling setup}\label{sec:ssc}

We want to give a prescription to calculate the three point functions for the ${\cal N}=4 $ SYM at strong coupling within the proposal of \cite{BlargeN}. The idea is to exploit the relation between the Gaussian matrix model and the OPE coefficient.

We should notice that the formulae for computing overlaps in equations (\ref{eq:gaussian1},\ref{eq:gaussian2}) is very similar to the formulae for computing averages in the thermal gas in equation (\ref{eq:vev}). 

Indeed, the main setup of \cite{BlargeN} 
states that half BPS wavefunctions associated to operators $\tr(Z^n)$ are 
computed at strong coupling by taking the ground state wave function in diagonal variables and multiplying it by $\tr(Z^n)$ where $Z= X_1+i X_2$ is a complex diagonal matrix (or other such product of traces). Also, the gaussian matrix measure is associated to the ground state wave function for the s-wave of the scalar fields in the free field limit \cite{Btoy}, when the field theory is compactified on a $S^3$. 

This same interpretation for the measure \ref{eq:vev} is available. Thus it is natural to compare the vacuum expectation values of equations  (\ref{eq:vev}) and (\ref{eq:gaussian1}, \ref{eq:gaussian2}).

Under this comparisson the 
same type of formula as above holds, where we substitute
\begin{equation}
\tr(Z^n) = \sum_p ( x^1_p+ix^2_p)^n \sim \int  z^n \rho
\end{equation}
This is, the traces are calculated in the matrix model of commuting matrices by summing over the eigenvalues. They can also be evaluated by calculating integrals over the density of particles in the Boltzman gas in the thermodynamics limit. Thus the integrals represent moments of the density distribution.

The density distribution of particles is given in the saddle point (thermodynamic limit) by \cite{BCV,BCott}
\begin{equation}
\rho_0(\vec x)=\vev{\rho(\vec x)} \sim \delta( |\vec x|- r_0)
\end{equation}
where $r_0= \sqrt{N/2}$, and the density is normalized so that $\int\rho= N$. 

We can think of the true density as given by $\rho= \rho_0+\delta \rho$, where $\delta \rho$
are density fluctuations and $\vev{\delta \rho}=0$. 

It is also convenient to normalize the size of the sphere so that the value of $r_0$ is scaled to one. This is done
by rescaling $\hat x =  \vec x/r_0$. In this way we find that
\begin{equation}
\tr(\hat z^n) = \int \rho(\hat x) \hat z^n \sim \int \rho(\hat x) Y_n(\theta) 
\end{equation}
This is, the normalized traces identify the integral of the density times a particular spherical harmonic of the sphere $S^5$. As such, these traces compute angular fluctuations of the
particle density on some specific spherical harmonics. We will use the notation $\delta\rho(n)$
to indicate schematically the density fluctuations integrated over the corresponding spherical harmonic, so that
\begin{equation}
\int \rho(\hat x) Y_n(\theta) \sim \delta\rho(n)
\end{equation}

The idea is to study the two point density fluctuations
\begin{equation}
\vev{\int \rho(\hat x) Y_n(\theta) \int \rho(\hat x) Y^*_n(\theta)}\sim \vev{\delta \rho^2(n)}
\end{equation}
and the three point density fluctuations
\begin{equation}
\vev{\int \rho(\hat x) Y_n(\theta) \int \rho(\hat x) Y_{m-n}(\theta)\int \rho(\hat x) Y_m^*(\theta)}\sim \vev{\delta \rho(n) \delta\rho(m-n)\delta\rho(m)}
\end{equation}
The first set of quadratic fluctuations can be considered the Gaussian fluctuations: given
the numbers $\vev{\delta^2\rho(n)}$ there is a unique Gaussian with central value zero whose  
fluctuations are characterized by those numbers.

The three point functions would vanish if the spectrum is fully Gaussian between the traces. Thus a non-vanishing of the three-point functions reflects a failure of the density distributions to be Gaussian. Since we expect the normalized three point functions to be of order $1/N$ (this follows from comparing to equation (\ref{eq:compare}), and from standard $1/N$ counting arguments \cite{'tH}), we expect the non-gaussianities to be small in the  large $N$ limit. The leading order non-gaussianities are these three point functions.

The method for calculating the three point functions is to generate a set of configurations distributed according to the measure \ref{eq:vev} via a Monte Carlo algorithm. The average
value of the two and three point functions can be calculated by averaging over a large sample of statistically independent typical configurations generated by the Monte-Carlo method. One can also sample the distribution of the $\delta\rho(n)$ directly and check that it is Gaussian.
The algorithm used was presented in \cite{BCott}. Here we use the same algorithm for various values of $N$. This will let us study the approach to the thermodynamic limit, to estimate the $1/N^2$ corrections. It is also useful to point out that a similar ensemble was studied in \cite{AN}, where a similar measure term was obtained. However, in that paper \cite{AN} there is no gaussian confining potential for the eigenvalues. This gaussian factor simplifies the analysis of the large $N$ limit considerably.

\section{Numerical results}\label{sec:nr}

We wish to study typical configurations of the gas of particles described by the equation (\ref{eq:vev}), and to evaluate the corresponding distributions for the possible values of various $\CO(x)$. We do this by studying individual configurations of the gas, with a 
measure proportional to
\begin{equation}
d\mu \sim \exp(-\sum_p \vec x_p^2+\sum_{i<j}\log|\vec x_i-\vec x_j|^2)
\end{equation}
The Monte-Carlo algorithm to navigate these configurations
 was described in \cite{BCott}. We move from one configuration to the next one by applying 
 a random move on particle $i$, where each of the coordinates of $\vec x_i$ is varied by a random number between $[-0.5\delta ,0.5\delta]$. We accept a new configuration according to the Metropolis-Hastings criterion. The typical values of $\delta$ that we use are between $4$ and $7$. We do this by cycling over all the particles in the gas and updating them one at a time. Roughly 40\% of the particles are updated in each cycle. 

We found that storing data from the configurations every 30 cycles was optimal. By this time the autocorrelations of the ensemble have died down considerably and for all practical purposes the 
configurations can be considered to be statistically independent. This is discussed in more detail in the appendix.

To obtain more information from each configuration, we chose various different orientations of the complex coordinate $z$, where 
\begin{equation}
z= x^\alpha+ix^\beta \hbox{ for }  \alpha<\beta
\end{equation} 
This choice is a symmetry of the system and it allows us to gain efficiency in the computation of the correlation functions. We wrote down only the multipoles themselves and not the full configuration of particles. 

For two point functions, we did the calculations with $N=400, 1K=10^3,  2K, 4K, 10K, 25K$ particles.
For three point functions, which require more statistically independent configurations to measure numbers statistically different than zero, we were able to get as high as 5K particles. We studied the first ten multipole coefficients $\delta\rho(i)$, for $i=1, \dots 10$.


\subsection{Two point functions}

The first measurements we made on the sample where those of the two point functions. These are expected to have a Gaussian distribution for $\delta\rho(n) \sim \int \delta\rho Y_n(\theta)$. The reason for this is that the
large $N$ counting responsible for equation \ref{eq:gauss} is robust independent of the the value of the 't Hooft coupling. We expect to find a Gaussian distribution for each $\delta\rho(n)$ in both the real and imaginary part. Because of the invariance of the ensemble under rotations of $z$ by phases, the real and imaginary part distribution will be the same. Also, the mean of the distribution vanishes.

\begin{figure}[ht]
\epsfxsize=8cm
\epsfbox{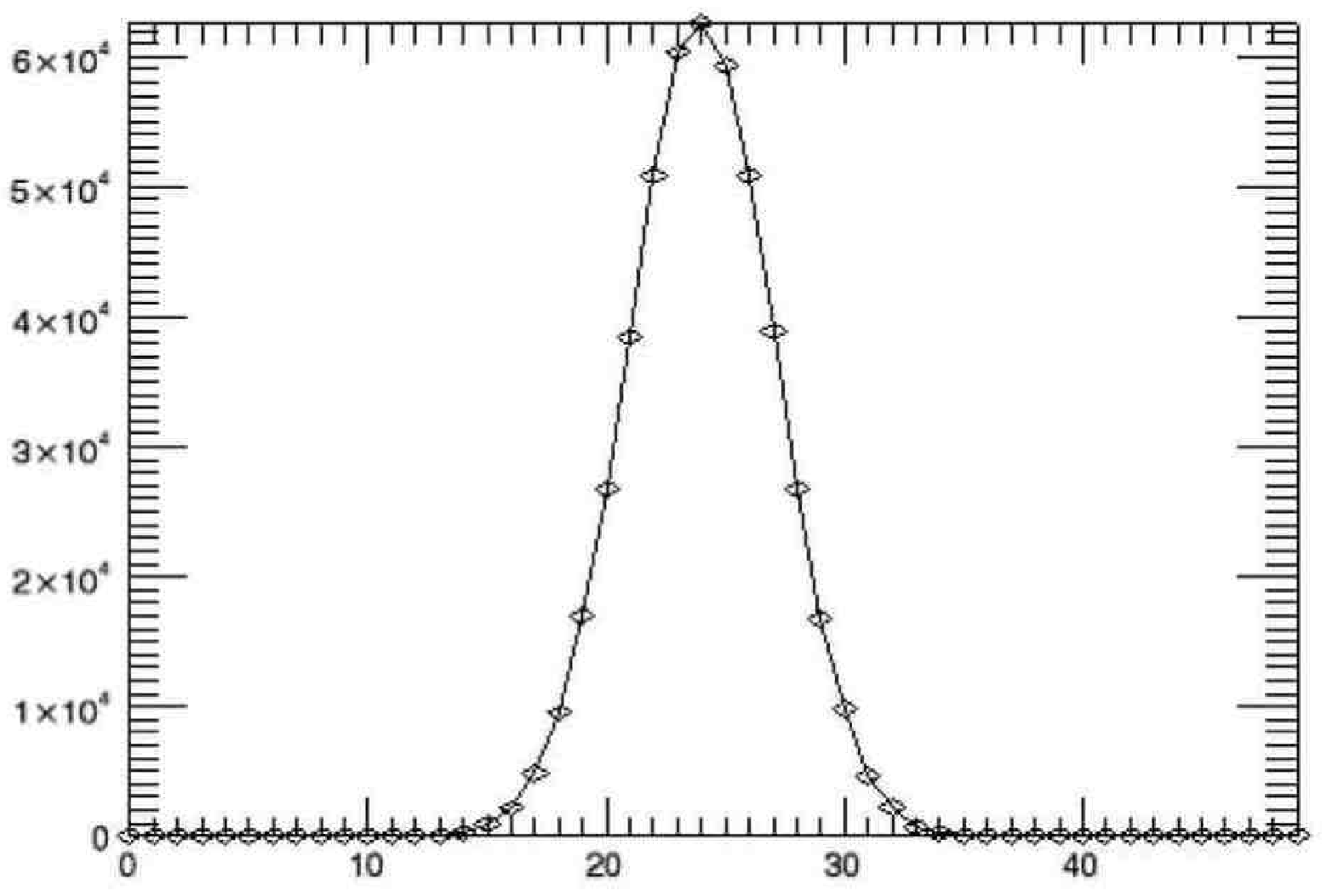} \epsfxsize=8cm \epsfbox{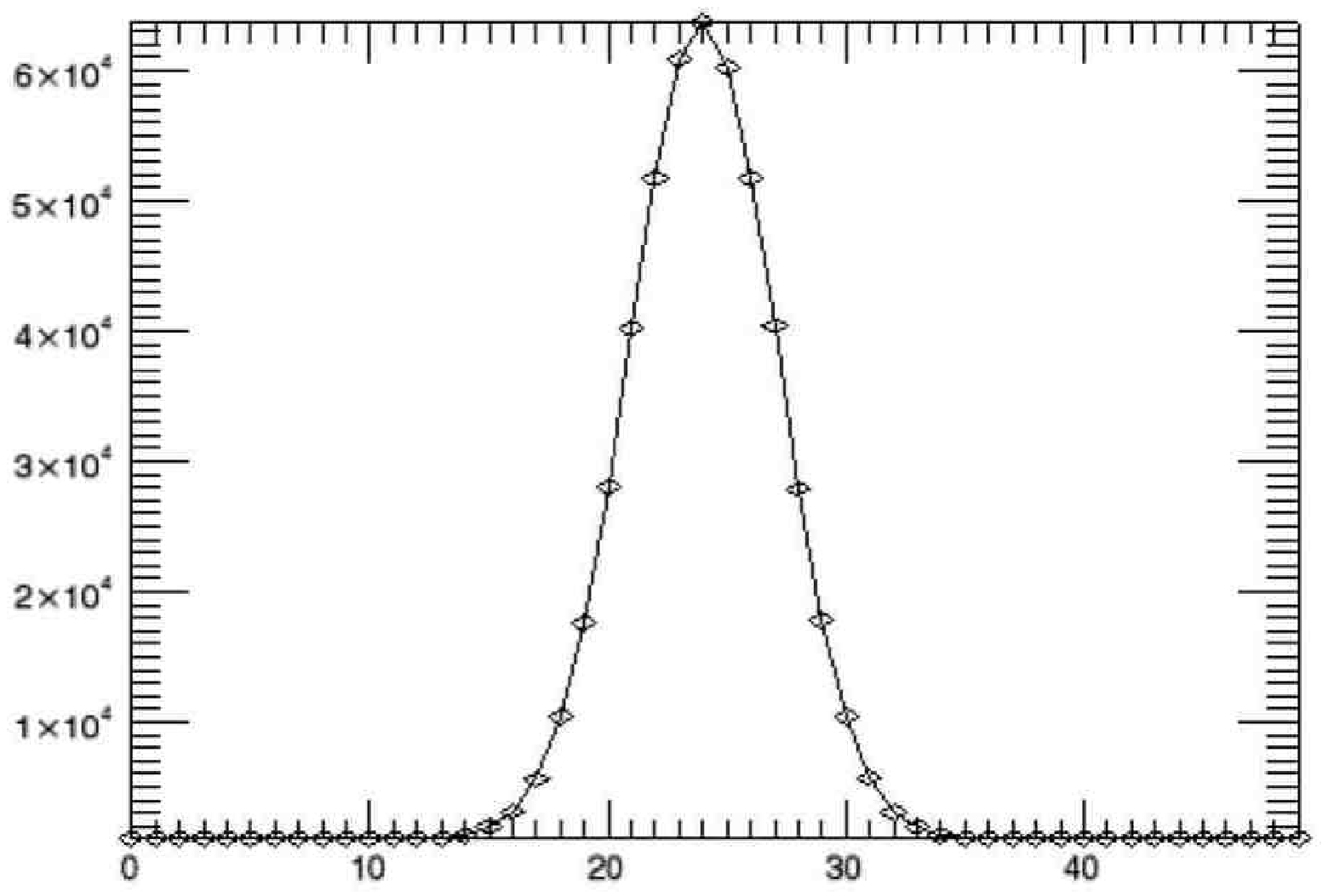}
\caption{Sample histogram distributions for $\delta\rho(n)$, N=5K. $n=1,2$}\label{fig:twopoint}
\end{figure}

To test for Gaussianity, we plot a histogram of the distribution in the interval $[-8\sigma, 8 \sigma]$ for the combined real part and imaginary part, divided into fifty intervals. This is depicted in figure \ref{fig:twopoint} for our largest statistical sample.

 One can put all ten multipole histograms on top of eah other, and they are statistically indistinguishable
 from one another. One also finds that the data is very well described by a Gaussian distribution.
 Indeed, it must be the case that the distribution for $\delta\rho(1)$ is Gaussian. This observation follows from writing the partition sum in terms of a center of mass variable 
 $\vec x_{CM} = N^{-1}\sum_p \vec x_p$, and relative coordinates. The two particle potentials are independent of $\vec x_{CM}$. The quadratic function
 $\sum_p \vec x_p^2$ can be written as a sum of squares of relative distances plus 
 a quadratic part. Thus the distribution for $\delta\rho(1)$ is Gaussian. This is also familiar from the full field theory for $U(N)$: the diagonal $U(1)$ degrees of freedom decouple completely.

If we normalize the radius of the distribution to one, we find that the fluctuations in the center of mass coordinate have a theoretical Standard deviation identically equal to one. We use this as a consistency check on the performance of the code. The fact that the histograms for all multipole moments look identical means we have a high degree of confidence that all the distributions are gaussian.

\begin{figure}[ht]
\epsfxsize=10cm
\epsfbox{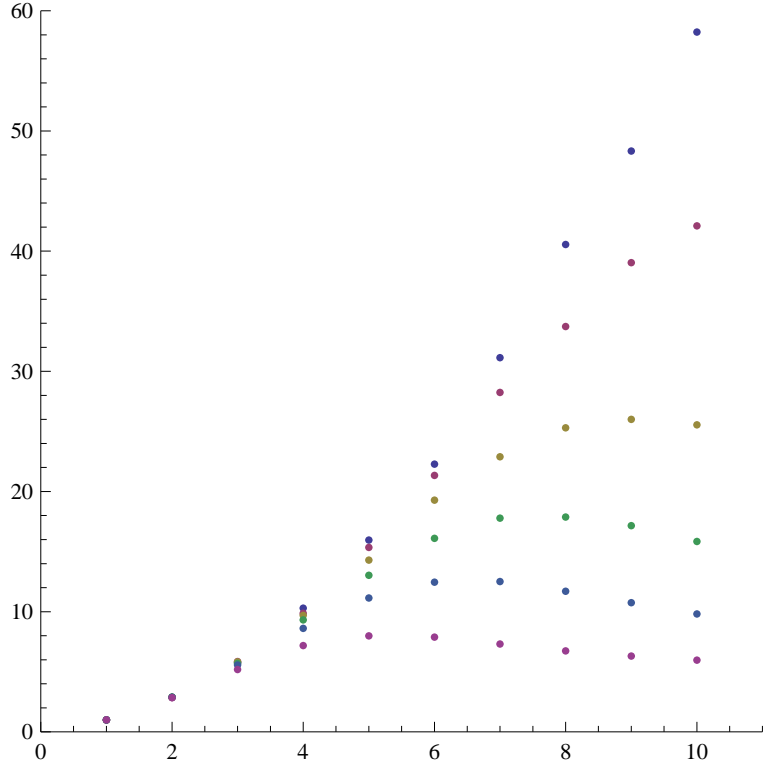}
\caption{Two point functions for
N=25K, N=10K, N=4K, N=2K, N=1K, N=400 respectively, from top to bottom. Labeled by angular momentum label $n$}\label{fig:twopointall}
\end{figure}

We can also look at the two point function absolute normalization, and how it depends on $N$.
The results we find are depicted in figure \ref{fig:twopointall}. Statistical error bars are 
small and are not shown. It is computationally cheap to generate large data samples to evaluate these precisely. Here we are interested in the qualitative approach to the large $N$ limit, not on measuring deviations from the large $N$ limit precisely (we expect that these should be a series in $1/N^2$ and are interesting on their own right, as they measure some type of quantum corrections in the dual theory, but their study is beyond the scope of the present paper).
 
The results for the absolute normalization of the two point functions are surprising in that the approach to the large $N$ limit for the individual multipole coefficients is slow. One can be off by 20\% in the normalization of $\delta(10)^2$ for $N=10^4$. This would suggest that $1/N^2$ corrections are rather large, even though $N$ is very large.

Indeed, the 
pattern seen in the figure \ref{fig:twopointall} suggests that the two point functions 
have a local maximum that depends on $N$. This local maximum moves to higher multipole 
moments as we increase $N$. Moreover, the maximum seems to occur at about half the
large $N$ limit value for the corresponding multipole moment (in our simulation the large $N$ limit is declared to be $N=25K$), so it can serve as a proxy for the scaling of the $1/N$ corrections. We find that 
the value of $J$ where the local maximum of $\delta(J)^2$ occurs seems to scale
like $N^{1/4}$ (the best fit to the data gives us $J\sim N^{1/3.9}$, but the value of $3.9$
has moderate systematic errors because $J$ can only take discrete values and we are not finding sufficiently large values of $J$ where these errors could become negligible). This is, corrections are suppressed by powers of ${J}^4/N$ (or equivalently $J/N^{1/4}$. The value of 
$N^{1/4}$ found here matches nicely with the expectation from gravity. After all, the radius of the sphere in $AdS_5\times S^5$  in the dual gravitational theory scales like $N^{1/4}$ in
Planck units. 

What this means is that corrections in $J$ (which is interpreted as momentum on the sphere) are suppressed
by the Planck scale. This is in essence another measurement of the Planck scale than the one found in \cite{BCott}, which was related to the back-reaction of the geometry to the presence of extended objects. This type of coincidences make it very plausible to believe that the matrix model of commuting matrices found in \cite{BlargeN} is actually capturing all the relevant physics of the strong coupling system that is described by a dual geometry. This is in contrast to the weak coupling (free field theory regime), where corrections scale like powers of $J^2/N$ \cite{allbmn}.

We should also contrast these results with the leading order two point function, as expressed in equation (\ref{eq:twoptfree}), where it is seen that the two point function scales as $C_{k,k}\sim k$, once the factors of $N$ are removed by the normalization of the $Z$.
Here we find that at strong coupling the value of this quantity scales differently with $k$. Indeed, the figure \ref{fig:twopointall} suggests that the scaling is power-law. A fit to the data suggests that
\begin{equation}
C_{k,k} \sim k^{1.75}
\end{equation}
more precisely, we make a fit for the power-law for various values of $N$, finding
the following values, depicted in table \ref{tab:exp}

\begin{center}
\begin{table}[ht]
\begin{tabular}{|c|c|c|c|c|c|c|c|}
\hline
$N$ &  400 & 1K & 2K & 4K & 19K &25K\\
\hline Exponent &
1.3&
 1.48&
 1.57&
 1.61&
 1.70&
1.74\\
\hline
\end{tabular}
\caption{Fit to power law behavior $C_{k,k}\sim k^\alpha$ for different values of $N$}\label{tab:exp}
\end{table}
\end{center}

The exponent $1.75$ is guessed because it is a simple rational number close to the 
observed value 1.74 $1.75= 2-1/4$. The
factor of $1/4$ in the denominator is also suggestive from the fact that the sphere scales like $N^{1/4}$ in Planck units in the gravity dual. It would be interesting if one can prove mathematically that the matrix model of commuting matrices predicts this value (this is, one would partially solve the matrix model).

\subsection{Three point functions}

Now we turn to the problem of evaluating three point functions numerically.
What we need to do is compute the average values of 
\begin{equation}
\vev{\tr(\hat z^n) \tr(\hat z^{m-n})\tr(\hat{\bar z}^{m})} \sim O(1/N) 
\end{equation}
and we expect the results to scale like $1/N$. For an individual configuration, the typical value of $\tr(\hat z^n)\sim \delta\rho(n)$ is found to be of order $1$. This means that we need to 
measure the center value of the  corresponding distribution by a factor of $1/N$ better than
the typical width of the distribution.

If we have a statistically independent sample of $L$ draws of the distribution, we find that 
the central value of the the distribution is measured to accuracy $1/\sqrt L$ relative to the width (this is the central limit theorem). We need this precision to be of order $1/N$ to distinguish the central value of the distribution from zero. This is, in order to measure 
a non-gaussian behavior in the distribution we need large statistical samples. In our case $L$ scales like $N^2$. We should also notice that in the Monte-Carlo code, each loop for updating the particles require $O(N^2)$ computations.
The CPU time required to get to the same precision in the measurement of non-gaussianities for higher values of $N$ scales like $N^4$.

\begin{figure}[ht]
\epsfxsize=14cm
\epsfbox{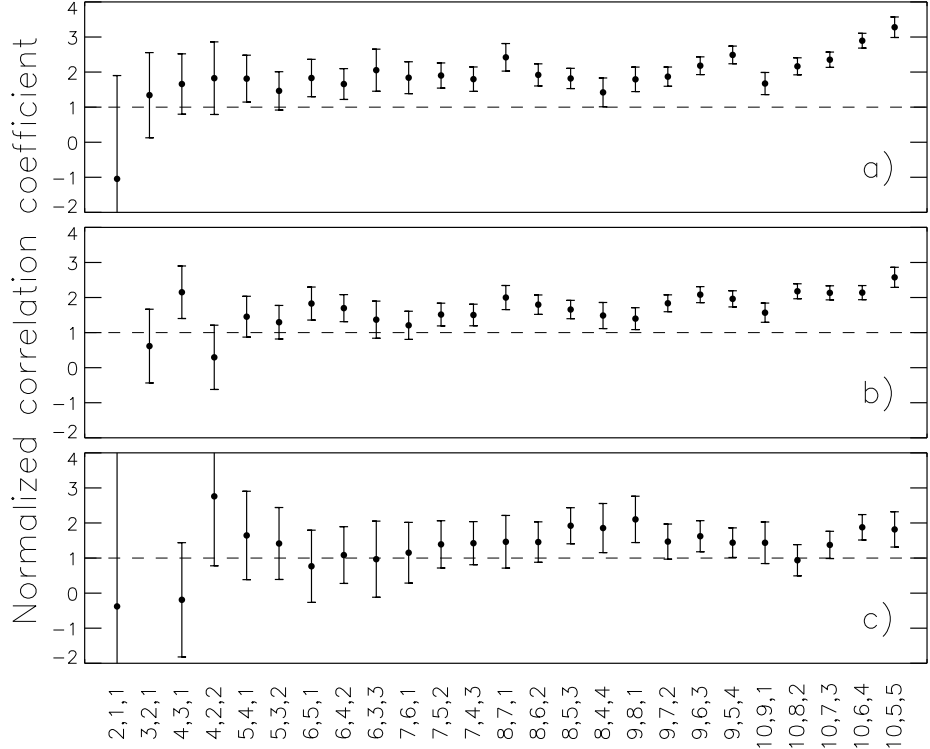}
\caption{Normalized three point functions. a) N= 1K, b) N= 2K, c) N= 5K, with $1\sigma$  error bars. The three integer labels in the $x$ axis are the values of $m,n,m-n$ multipoles that we are comparing.}\label{fig:3pt}
\end{figure}

The possibility of a non-renormalization theorem as described in \cite{LMRS} lets us 
compare with the values of these non-gaussianities in free field theory, as expressed by 
\ref{eq:compare}. This is, we can measure in the Monte-Carlo and compare absolutely
to the result found in gravity by rescaling the three point function we measure by the expected value. Thus, we can compare universally for various values of $N$ if we are near the expected value or not. Seeing as the two point functions behave differently at strong coupling than at weak coupling in figure \ref{fig:twopointall}, a match of the three point functions
would be a strong test that the matrix model of commuting matrices is correct.
Our results are presented in figure \ref{fig:3pt}.

The most important number for our considerations is $L$,  the number of statistically
independent configurations. We have that for $N= 5K$ we have a total of 242151
measurements with very small correlations. This is about a tenth of $(5000)^2$, which is the suggested number of configurations that we would need. This sample took over a month of CPU time to generate. What makes the calculation possible is that the coefficient we are trying to measure is of order $\sqrt{n(m-n)m}$ larger than just $1/N$. This number can be as large
as $15$, and is typically of order $7$, so for these values we can cut the number of
samples by $1/7^2$ to get reasonable results. Thus, with the number of samples we have it is possible to test the commuting matrix model.

The value that we are comparing to for each triplet of mutipoles is normalized to one. We plot the values of $N_{n,m-n;m}$ that are measured in the statistical sample with a $1\sigma$ error bar. We notice that the error bars become smaller as we increase $m$. 
This is expected because the corresponding three point functions are larger relative to the width of the distribution.

On the other hand, we expect that as we go to higher multipole moments the ``quantum corrections" become larger and larger and we are further away from the large $N$ limit 
value. We should also have in mind that the center of mass mode decouples completely (this is referred to as the singleton sector in the supegravity theory), so one should ignore the correlations with $n=1$ for the most part. They are artificial in that we have defined $\tr(\hat z^n)$ without substracting the center of mass mode (a discussion of these issues can also be found in \cite{deK}). Indeed, this is what one would do if one were studying the $SU(N)$ theory rather than $U(N)$. In general, this is a $1/N$ correction to the traces,
and these small corrections will be of order $1/N^2$ in the definition of the other three point functions. Thus we tabulate them in the unsubstracted form, which is computationally simpler, without making any significant errors. We left the data in the graph with $n=1$ to show all of
our numerical results.

We see from the figure \ref{fig:3pt} that as $N$ increases, the match to the theoretical supergravity value 
improves. We do not have a clear understanding of the systematic deviations from being at finite $N$ rather than in the thermodynamic limit (we will call this our systematic error). In practice, we would need larger values of $N$ and much larger statistical samples to be able to study this question numerically.

\begin{figure}[ht]
\epsfxsize=14cm
\epsfbox{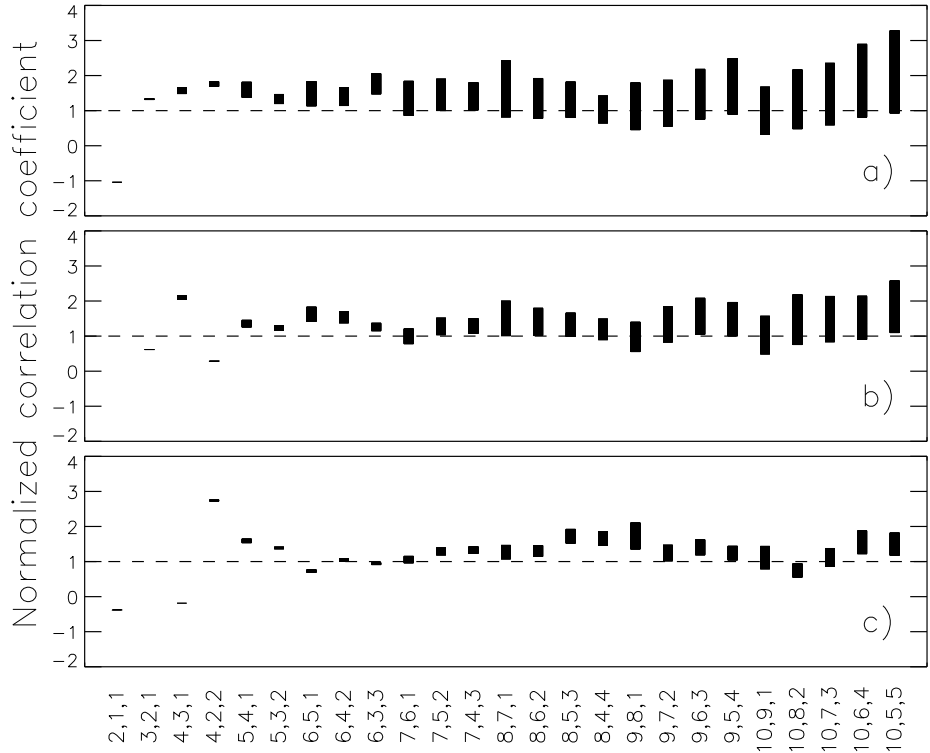}
\caption{A naive estimate of the systematic corrections for the same data of figure \ref{fig:3pt}.}\label{fig:3pterr}
\end{figure}

However, we can estimate the size of the expected variations by using the deviations of the two point functions as given in figure \ref{fig:twopointall} as a proxy for the systematic errors. This is, we take the non-normalized three point functions and divide them by the correct value of the two point functions in the thermodynamic limit (four our purposes this is $N=25K$. The deviations in the central value are depicted in figure \ref{fig:3pterr}. The large value corresponds to the measured two point functions at the given value of $N$ while the smalll value corresponds to 
the two point function normalization for large $N$. A better theoretical estimate would have us 
fit the data to an expansion in $1/N^2$ and extrapolate to $N\to \infty$. For this to work, we need to ensure that we are in a regime where only the $1/N^2$ correction matters. Unfortunately, the two point functions seem to indicate that higher corrections in $1/N^2$ 
are relevant. Also, our data quality is too poor to do this reliably and we consider the estimate described above more meaningful.

From the figure \ref{fig:3pterr} we see that a naive analysis of the systematic corrections suggests that the values will 
move closer to the gravity value, and that these are becoming smaller as $N$ grows.

\begin{figure}[ht]
\epsfxsize=16cm
\epsfbox{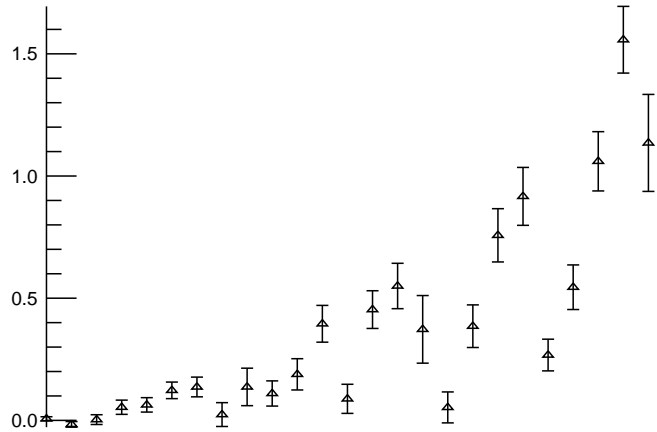}
\caption{Non-normalized data for three point functions for N=5000}\label{fig:rawdata}
\end{figure}

Just for comparison, we include here a graph for the non-normalized three point functions
in figure \ref{fig:rawdata}, for the particular case of $N=5K$. As the graph shows, the pattern of three point functions has a lot of variation in the numbers that we measure (the range is over two orders of magnitude). The fact that in figure \ref{fig:3pt} the data becomes very flat and of order one suggests very strongly that modeling the data by the gravity prediction gives a very reasonable fit to the numerical data.

Our final numerical results are that the matrix model calculation is consistent with the gravity computation. Since the systematic errors are still large one can not claim at this point that there is an exact match, but there is strong evidence for agreement. Considering that there were large corrections in the two point functions, we believe this is a strong test of the commuting matrix model proposal. 

\section{Conclusion}

In this paper we have studied the extremal three point functions in the commuting matrix model proposal for the strong coupling expansion of ${\cal N}=4 $ SYM at strong coupling \cite{BlargeN}. We did this by calculating a statistical average of the ground state wave function of the model, following a Monte Carlo algorithm that was used in the previous work  \cite{BCott}. Considering that the values of the two point functions that we measured were very different than what was expected from the weak coupling calculations (they have different scaling with the multipole number), it seems rather surprising that the normalized three point functions match. Although this is expected from non-renormalization theorems \cite{EHSSW}, it is hard to imagine that with all the approximations that are made to derive the model in \cite{BlargeN} that the normalized three point functions are essentially uncorrected.

We have also seen numerically that the $1/N$ corrections have different parametric dependence on the quantum numbers of a state than at weak 't Hooft coupling. This fact motivates us to try to understand better the proposal given in \cite{BlargeN}. 
Recent work in studying more general examples of the AdS/CFT correspondence \cite{Bhart} has shown that this type of proposal can be extended to many other setups. It has already been shown that a big part of the supergravity spectrum including many non-BPS multiplets is matched. That is exactly the collection of states that we are able to simulate in the${\cal N}=4 $ SYM setup.

Thus, in principle, if one is able to numerically evaluate the distributions of particles that
generate other geometries, one would be in a position to study three point functions of 
primary operators on strongly coupled field theories numerically. Here one can do a strong test of the AdS/CFT correspondence by matching the three point functions of field theory to those of gravity. Seeing as in many of these field theories there is no perturbative limit where some of these calculations could be performed, it would be the first instance of calculating the field theory correlators (and therefore the OPE expansion coefficients) directly at strong coupling.
To do this one needs some detailed information about the metric of Sasaki-Einstein manifolds, and it might be necessary to compute these numerically along the lines of  \cite{DHHKW}.

It would also be interesting to obtain better statistics and a systematic determination of how 
the $1/N$ corrections behave for both two and three point functions in ${\cal N}=4$ SYM. These encode non-trivial higher genus corrections in the dual string theory.

\section*{Acknowledgements}

D. B. would like to thank J. Nishimura, M. Rydenfelt, R. Sugar, T. Wiseman for many discussions. D.B. work supported in part by the U.S. Department of Energy, under grant DE-FG02-91ER40618. R.C. work supported by an NSF Graduate Fellowship. R. L. is supported by the NASA Planck project
under JPL contract \#1261740.

\appendix

\section{Statistical error bars and correlations}

In this appendix we give for completeness a brief description of the computation of the statistical error bars in our results, as well as how we optimized the extraction of data from 
the Monte Carlo generator. This is important because some of the measurements we are 
making (the calculation of three point functions) require calculating an average of a quantity to  precisions much higher than it's typical fluctuation. Indeed, we are trying to measure quantities of order $1/N$ to fluctuations of order one.

The main observation is that when we compute an average in a Monte Carlo setup, we are computing the average of a time series, where the time in question is simulation time.

Thus, we are computing
\begin{equation}
\vev{h} = \frac 1 L \sum_{t=1}^{L} h(t)  \label{eq:av}
\end{equation}
where $L$ is the number of recorded loops.

We want to estimate the statistical error bar for $h$. If the $h(t)$ are statistically independent, then the variances satisfy
\begin{equation}
\sigma^2( \vev{h}) \sim \sigma^2(h)/L
\end{equation}
 
 However, in practice, the $h(t)$ are correlated. The correlations can be computed by the covariance of the time series
 \begin{equation}
 c(t) \sim \frac{\vev{ \delta h(t+i) \delta h(i)}}{\vev{\delta h^2}} 
 \end{equation}
 
 Using $c(t)$, one can compute the error in equation (\ref{eq:av}) by using the standard formula (see \cite{Liu}, eq. 5.13,  for example)
 \begin{equation}
 \sigma^2( \vev{h}) \sim L^{-1}\sigma^2(h)(1+2\sum_{t=1}^n c(t))
 \end{equation}
where $n$ is a suitable cutoff.

Typically the $c(t)$ decay exponentially, $c(t) \sim \exp(-\lambda t)$, where $\lambda$ depends on the observable. To some extent $\lambda^{-1}$ measures the relaxation
time of the corresponding variable, so it is  convenient to plot $\log(|c(t)|)$ with respect to 
simulation time for various variables. 

\begin{figure}[ht]
\epsfxsize=13cm
\epsfbox{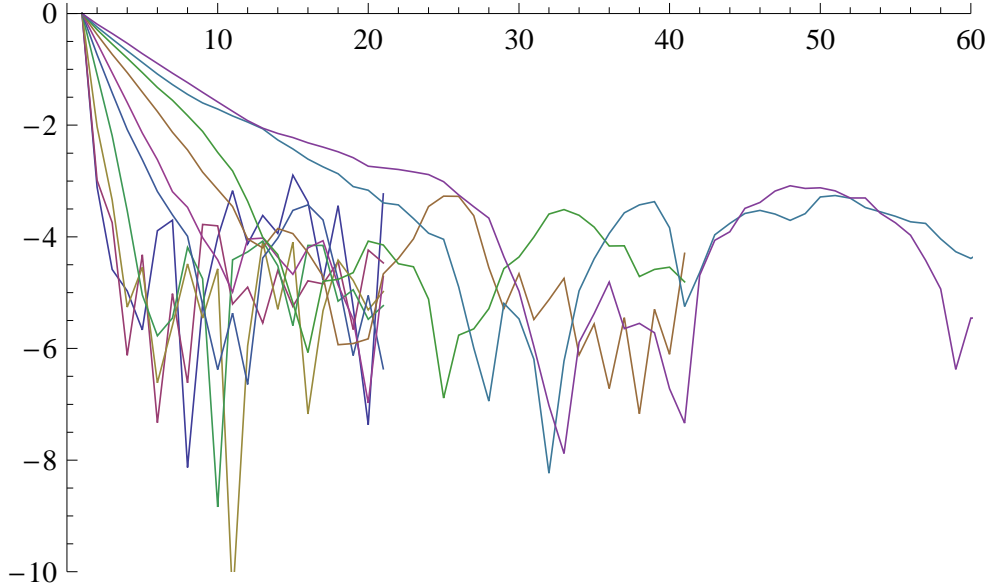}
\caption{Logarithm of Correlations of two point functions, for N=25K, steps per recording=5}\label{fig:corr}
\end{figure}

The figure \ref{fig:corr} shows the logarithm of the autocorrelations of the different multipoles for the simulation with $N=25\times 10^3$ particles, and recording the
data after five simulation steps. The different curves represent the different multipoles.
It is clearly seen that at the beginning there is a typical linear behavior, with different values of $\lambda$ for each multipole.
 There is usually a value where $c(t)$ becomes oscillatory. This is a reflection of the limited 
statistical sample and it is a good place to cutoff the value of $n$ used in the sum above.

For the data shown in figure \ref{fig:3pt} the step used between recordings was 30 Monte Carlo iterations, where the two point autocorrelations decay very quickly (six times faster than in figure \ref{fig:corr}). We tuned the steps for that simulation so that roughly $log(|c(1)|)<-1$ for all two point function observables when we are doing three point function measurements. The statistical autocorrelation was used to write the error bars in figure \ref{fig:3pt}. This resulted in a correction of less than about 
3\% for all values, if one assumed that the all the data taken was statistically independent.

 Although individual values of $\lambda$ depend on how one generates configurations, the ratios $\lambda_i/\lambda_j$ might show a more stable pattern
when one changes the parameters of a simulation.
In our case we see that the
highest multipoles have longer autocorrelations. If one considers the $\lambda^{-1}$ numerically, they seem to show a pattern that scales simply with the mutipole number. This might be interesting to explore for the future.

\end{document}